\documentclass{aa}
\usepackage{psfig}

\begin{document}

   \thesaurus{06         % A&A Section 6: Form. struct. and evolut. of stars
              (08.02.3;  % Stars:binaries:close,
               08.14.2;  % Stars:neutron,
               08.09.02 GRS 1741.9$-$2853;  % Stars:individual,
               13.25.1;  % X-rays:bursts,
               13.25.5)}  % X-rays:stars
   \title{Discovery of type-I X-ray bursts from GRS~1741.9--2853}
%   \subtitle{}

   \author{ M. Cocchi \inst{1}, A. Bazzano \inst{1}, L.Natalucci \inst{1}, P. Ubertini \inst{1},
            J. Heise \inst{2}, J.M. Muller \inst{2,3} and J.J.M in 't Zand \inst{2}
          }
   \offprints{M. Cocchi, wood@ias.rm.cnr.it}

   \institute{$^{1}$ Istituto di Astrofisica Spaziale ({\em IAS/CNR}), via Fosso del Cavaliere 100, 00133 Roma, Italy \\
              $^{2}$ Space Research Organization Netherlands {\em (SRON)}, Sorbonnelaan 2, 3584 TA Utrecht, The Netherlands \\
              $^{3}$ {\em BeppoSAX} Science Data Centre, Nuova Telespazio, via Corcolle 19, 00131 Roma, Italy
             }

   \date{Received ?; accepted ?}

   \authorrunning{Cocchi et al.}
   \maketitle

   \begin{abstract}
For the first time X-ray bursts have been detected from a sky position 
consistent with the one of GRS 1741.9$-$2853, a {\em GRANAT} transient 
source located only $\sim 10\arcmin$ from the Galactic Centre.
A total of 3 bursts have been observed by the Wide Field Cameras telescopes
on board {\em BeppoSAX} during a monitoring observation of the Sgr~A region
in August-September 1996. The characteristics of the events are consistent 
with type-I bursts, thus identifying the source as a likely low-mass
X-ray binary containing a neutron star.  Evidence of photospheric radius
expansion due to super-Eddington luminosity is present in one of the 
observed bursts, thus leading to an estimate of the source distance 
($\sim 8$ kpc).

      \keywords{binaries:close -- stars: neutron, individual (GRS 1741.9$-$2853) -- X-rays: bursts}
   \end{abstract}

%--------------------------------------

\section{Introduction}

GRS 1741.9$-$2853 was discovered during the first observations of the Galactic
Centre region performed by the {\em GRANAT} satellite in Spring 1990.
The source was detected by the low-energy (4--30 keV) imaging telescope ART-P
in the March~24--April~8 observations and was tentatively associated (\cite{Mand90}) 
to the soft {\em EINSTEIN} source 1E 1741.7$-$2850 (\cite{Wats81}).
Further analysis of the same data (\cite{Suny90,Suny91,Pavl94}) refined the
source position, obtaining $\alpha=17^{\rm h}41^{\rm m}50^{\rm s}$, 
$\delta=-28^{\circ}52^{\prime}54\arcsec$ (B1950, error radius $45\arcsec$, 
90\% confidence). 
The possible association with 1E 1741.7$-$2850 and also with the nearby {\em GINGA} 
transient GS 1741.2$-$2859/1741.6$-$289 (\cite{Mits90}) was ruled out.
The average 4--20 keV intensity of GRS 1741.9$-$2853 was $9.6\pm0.7$ mCrab, 
corresponding to $(1.6\pm0.1)\times10^{36}{\rm erg~s}^{-1}$ at 8.5 kpc,
and the source spectrum could be fitted by a thermal bremsstrahlung with a 
temperature of $\sim8$ keV.
During the same {\em GRANAT} observations, GRS 1741.9$-$2853 was detected above the
$3.5\sigma$ detection level of the soft Gamma-ray telescope SIGMA in the 40--100 keV
band (\cite{Chur93,Varg97}).
On the other hand, both ART-P and SIGMA did not detect the source $\sim 4$ months 
later during the Fall 1990 observation campaign, suggesting GRS 1741.9$-$2853 
is transient in nature.
A $3\sigma$ upper limit of 1.2 mCrab was obtained by ART-P in the 4--20 keV band
(\cite{Pavl94}), thus implying a drop in the source intensity of at least a factor 
of 7.  Moreover, {\em GRANAT} failed to detect GRS 1741.9$-$2853 in all the subsequent 
campaigns on the Galactic Centre (Spring 1991, Fall 1991, Spring 1992, e.g. 
\cite{Chur93,Pavl94,Varg97}).
The source was not observed in detailed mappings of the Galactic Centre region by 
soft X-ray instruments like {\em EINSTEIN} (0.5--4.5 keV, $3\sigma$ upper limit of 
$\sim0.7$ mCrab, see \cite{Wats81}), {\em SPACELAB-2} (3--30 keV, $<0.8$ mCrab, \cite{Skin87}), 
and by {\em ROSAT} (0.8--2.5 keV, $<0.1$ mCrab, \cite{Pred94}), thus confirming its 
transient nature.
More recently, no detections of GRS 1741.9$-$2853 were reported by {\em RXTE}-ASM in 
the 2--10 keV energy band since February 1996.

In the next section we briefly introduce the Wide Field Cameras telescopes and
report on the observations of GRS 1741.9$-$2853.  Time resolved spectroscopy
of the burst data is presented in Section 3, while the impact
of our results on the knowledge of the source are discussed in Section 4.
In particular, we propose GRS 1741.9$-$2853 as a transient low-mass X-ray binary 
harbouring a neutron star and we give an estimate of the source distance.

%----------------------------------------------------------- Figure 1
    \begin{figure}[hbt]
      \psfig{figure=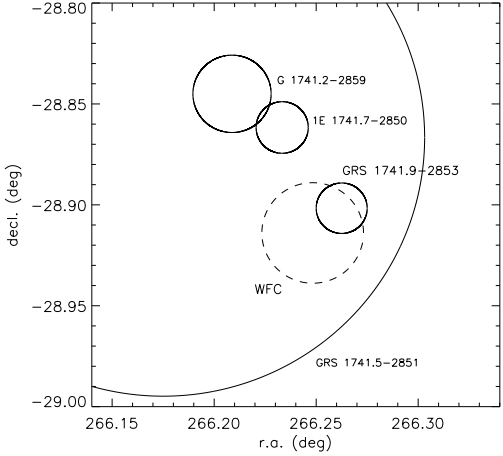,width=7cm,clip=t}
      \caption{
               A plot of source error circles (J2000, 90\% confidence) in the 
               GRS 1741.9$-$2853 sky region.  The dashed circle refers to the WFC 
               observed bursting source.
              }
      \label{Fig1}
   \end{figure}
%______________________________________________________________

%------------------------------------------------------------------- Table 1
\begin{table}[hbt]
\caption{Summary of the characteristics of the observed bursts
}
\protect\label{t:ae}
\begin{flushleft}
\begin{tabular}{lccc}
\hline
\hline \noalign{\smallskip}
parameter                             &  Burst 1 & Burst 2 & Burst 3 \\
\hline \noalign{\smallskip}
Burst date                            & Aug. 22 & Aug. 24 & Sep. 16 \\
Burst UT time (h)                     & 23.4130 & 3.7972 & 9.6103 \\
{\em e}-folding time (s)              & $\ga 8.8\pm 2.5$  &  $11.0\pm 1.7$  & $16.0\pm 1.5$ \\
peak intensity $^{(a)}$               & $\ga 384\pm 62$  &  $547\pm 56$  & $983\pm 69$  \\
{\em k}T (keV)                        & $1.78\pm 0.16$ & $2.26^{+0.14}_{-0.13}$ & $1.94\pm 0.07$ \\
Reduced $\chi^{2~(b)}$                & 1.23 &  1.15   &  0.95  \\
$R_{\rm km}/d_{10~{\rm kpc}}$         & $7.9^{+2.1}_{-1.4}$ & $6.3^{+1.0}_{-0.8}$ & $10.4^{+0.9}_{-0.8}$ \\
$N_{\rm H}$ ($10^{22} {\rm cm}^{-2}$) & $10.2^{+10.1}_{-9.7}$  &  $36^{+21}_{-13}$  &  $10.3^{+3.7}_{-2.7}$ \\
steady emission $^{(c)}$              & 3 & 3 & 10 \\
\noalign{\smallskip}
\hline\noalign{\smallskip}
\noalign{$^{(a)}$ in mCrab, 3-28 keV band; $^{(b)}$ 26 d.o.f.;
 $^{(c)}$ $3\sigma$ upper limits, in mCrab, 2-10 keV.} 
\end{tabular}
\end{flushleft}
\end{table}
%_______________________________________________________________________

\section{Observations}

One of the main scientific objectives of the Wide Field Cameras (WFC) on board 
the {\em BeppoSAX} satellite is the study of the timing/spectral behavior of both 
transient and persistent sources of the Galactic Bulge region, X-ray binaries 
in particular, on time scales from seconds to years.  To this end,
an observation program of systematic monitoring of the Sgr~A sky region is being 
carried out (e.g. \cite{Heis98}).
The WFCs consist of two identical coded mask telescopes (\cite{Jage97}) pointing in 
opposite directions.
Each camera covers a $40\degr \times 40\degr$ field of view, the largest ever flown 
for an arcminute resolution X-ray imaging device.
With their source location accuracy in the range $1\arcmin$--$3\arcmin$ (99\% confidence),
a time resolution of 0.244 ms at best, and an energy resolution of 18\% at 6 keV, the
WFCs are very effective in studying X-ray transient phenomena in the
2--28 keV bandpass.  The imaging capability and the good instrument sensitivity
(5-10 mCrab in $10^{4}$ s) allow an accurate monitoring of complex sky regions,
like the Galactic bulge.
The data of the two cameras are systematically searched for bursts and flares by 
analyzing the time profiles of the detectors in the 2--11 keV 
energy range with a time resolution down to 1 s.  Reconstructed sky images are
generated for any statistically meaningful event, to
identify possible bursters. The accuracy of the reconstructed position, which 
of course depends on the burst intensity, is typically better than $5\arcmin$.
This analysis procedure demonstrated its effectiveness throughout the Galactic Bulge
WFC monitoring campaigns (e.g. \cite{Cocc98a}), leading to the identification of 
$\sim 700$ X-ray bursts (156 of which from the {\em Bursting Pulsar} GRO J1744$-$28) in 
a total of about $2\times10^{6}$ s net observing time.  A total of 13 new X-ray bursting sources 
were found, thus enlarging the population of the bursters by $\sim 30\%$ (\cite{Heis99,Uber99}).

GRS 1741.9$-$2853 is in the field of view whenever the WFCs point at 
the Galactic Centre region, being only $\sim 10\arcmin$ away from the Sgr A position. 
No steady emission was observed during the whole WFC monitoring campaign. 
Typical 2--10 keV $3\sigma$ upper limits of $\sim 3$ mCrab were derived (see Table 1).

Three X-ray bursts were detected at a position consistent with that of GRS 1741.9$-$2853
in two different observations (Aug. 21.774--31.519 and Sep. 13.408--18.254) during the Fall 
1996 monitoring campaign.
Due to the {\em BeppoSAX} orbit characteristics, the source covering efficiency during an 
observation is in average $\sim 53\%$, so other bursts could be missed.
The averaged error circle obtained for the position of the bursting source is shown 
in Fig. 1.  None of the observed bursts can be associated to other known sources.
In Fig. 2 the time profiles of the three bursts are displayed.  The August 22 burst 
occurred in coincidence with a $\sim 10$ s telemetry gap and some seconds of data belonging 
to the leading part of the burst are missed. So we can not determine the burst on-time with 
sufficient accuracy.  The characteristics of the observed bursts are summarized in
Table 1.
An accurate search for bursts from GRS 1741.9$-$2853 was performed on all the data available from 
the 1996-1998 {\em BeppoSAX}-WFC Galactic Bulge monitoring campaigns but no other events 
were observed.

%----------------------------------------------------------- Figure 2
    \begin{figure}[ht]
      \psfig{figure=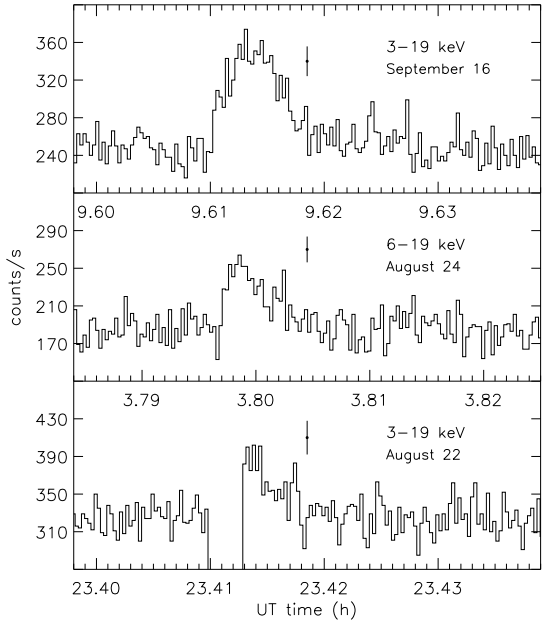,width=8cm,clip=t}
      \caption{
               Time profiles of the three observed bursts.  Energy ranges 
               maximizing the S/N ratio were selected.
              }
         \label{Fig2}
   \end{figure}
%______________________________________________________________

%----------------------------------------------------------------------- Table2
\begin{table}[hb]
\caption{ Time resolved spectral analysis of bursts 2 and 3
}
\protect\label{t:hc}
\begin{flushleft}
\begin{tabular}{lccc}
\hline
\hline\noalign{\smallskip} 
time range$^{(a)}$ & {\em k}T (keV) & $R_{\rm km}/d_{10~{\rm kpc}}$  & $\chi^{2~(b)}_{r}$ \\ 
\noalign{\smallskip} \hline\noalign{\smallskip} 
\multicolumn{4}{c}{{\bf August 24 burst}} \\
 $T_{0} \div T_{0}+5s$      & $1.96^{+0.21}_{-0.18}$ & $9.0^{+2.8}_{-1.8}$ & 1.09  \\
 $T_{0}+5s \div T_{0}+11s$  & $2.70^{+0.25}_{-0.22}$ & $4.9^{+1.2}_{-0.8}$ & 0.96  \\
 $T_{0}+11s \div T_{0}+20s$ & $2.11^{+0.27}_{-0.22}$ & $6.3^{+2.3}_{-1.4}$ & 0.72  \\
\noalign{\smallskip}
 \multicolumn{4}{c}{{\bf September 16 burst}} \\
 $T_{0} \div T_{0}+3s$      & $1.96^{+0.33}_{-0.29}$ & $8.7^{+4.3}_{-2.2}$ & 0.95  \\
 $T_{0}+3s \div T_{0}+15s$  & $1.71^{+0.09}_{-0.08}$ & $14.8^{+1.8}_{-1.5}$ & 1.34 \\
 $T_{0}+15s \div T_{0}+20s$ & $2.62^{+0.17}_{-0.16}$ & $6.7^{+1.0}_{-0.8}$ & 0.81  \\
 $T_{0}+20s \div T_{0}+30s$ & $1.82^{+0.17}_{-0.16}$ & $9.3^{+2.3}_{-1.6}$ & 1.01  \\
\noalign{\smallskip} \hline\noalign{\smallskip}
\noalign{ $^{(a)} T_{0}$ indicates the burst time (see Table 1);
          $^{(b)}$ 27 d.o.f.}
%          $^{(c)} N_{\rm H}$ frozen to $36.0 \times 10^{22}~{\rm cm}^{-2}$;
%          $^{(d)} N_{\rm H}$ frozen to $10.3 \times 10^{22}~{\rm cm}^{-2}$. }
\end{tabular}
\end{flushleft}
\end{table}
%_____________________________________________________________________________________

\section{Data Analysis}

Energy resolved time analysis of the bursts was performed to study the spectral 
evolution of the observed events.
Due to the above mentioned missing data in the August 22 burst observation, only 
the August 24 and September 16 bursts were analyzed this way (see Fig. 3).
The time histories of the bursts are constructed by accumulating only 
the detector counts associated with the shadowgram obtained for the sky position of the 
analyzed source, thus improving the signal to noise ratio of the profile.  
The background is the sum of (part of) the diffuse X-ray background, the particles 
background and the contamination of other sources in the field of view.
Source contamination is the dominating background component for crowded sky fields like 
the Galactic Bulge.  Nevertheless, the probability of source confusion during a short 
time-scale event (10--100 s) like an X-ray burst is negligible.

The burst spectra of GRS 1741.9$-$2853 are consistent with absorbed blackbody 
radiation with average color temperatures of $\sim 2$ keV. 
A summary of the spectral parameters of the three bursts is given in Table 1.
The value of the $N_{\rm H}$ parameter obtained for the August 24 burst is higher 
with respect to the August 22 and September 16 ones.  
For burst 2, freezing the $N_{\rm H}$ value to the average value of burst 1 and 3 
($10.3\times10^{22} {\rm cm}^{-2}$) leads to higher values of the reduced $\chi^{2}$ 
(1.30 for 27 d.o.f.), to an higher blackbody 
temperature ($2.64\pm 0.15$ keV) and to a lower blackbody radius ($3.9\pm 0.4$ 
km at 10 kpc).  Conversely, if we assume that all the three bursts had 
in average the same characteristics (color temperature and radius of the emitting sphere),
this implies a 1-day time-scale $N_{\rm H}$ variability of a factor of $\sim 3$.

Time-resolved spectra were accumulated for burst 2 and 3, in order to study the time
evolution of their spectral parameters.
To better constrain the fits, the $N_{\rm H}$ parameter was kept fixed, according to 
the values obtained for the total bursts, i.e. $36.0\times 10^{22}{\rm cm}^{-2}$ and 
$10.3\times 10^{22}{\rm cm}^{-2}$
for burst 2 and burst 3 respectively.
Blackbody spectra allow to determine the relationship between the average radius of the 
emitting sphere $R_{\rm km}$ (in units of km) and the source distance $d_{\rm 10~kpc}$ 
(in units of 10 kpc).
In Fig. 3 and in Table 2 the time histories of the measured $R_{\rm km}$/$d_{\rm 10~kpc}$ ratios 
are shown, assuming isotropic emission and not correcting for gravitational redshift and 
conversion to true blackbody temperature from color temperature (see \cite{Lewi93} 
for details). A radius expansion of a factor of $\sim 2$ is observed in the September 16 burst.

%----------------------------------------------------------- Figure 3
    \begin{figure*}[htb]
      \psfig{figure=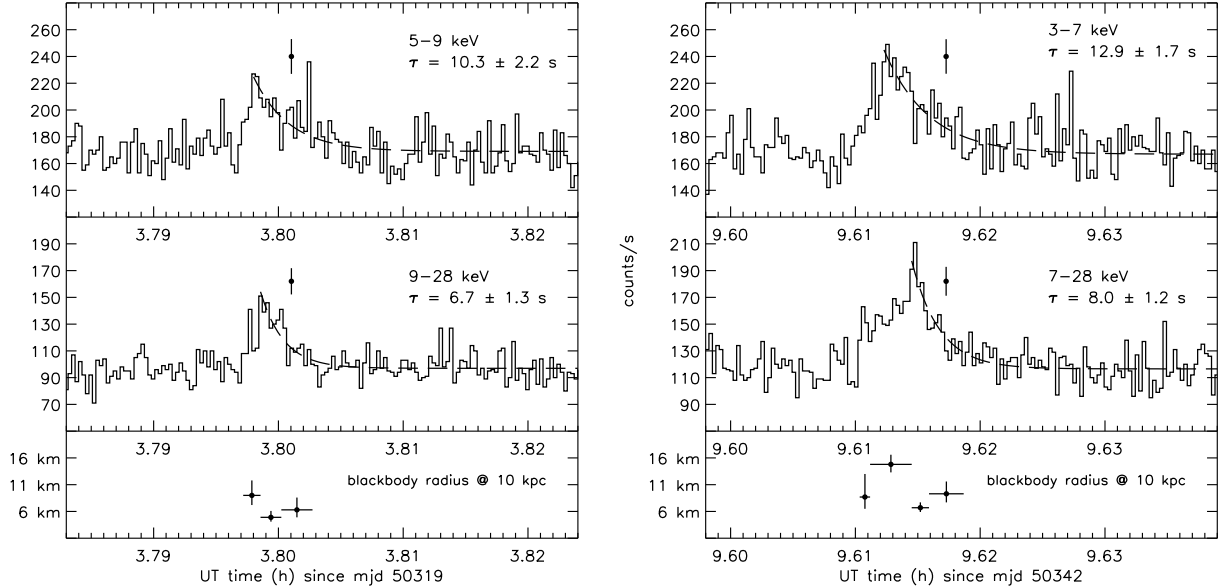,width=17cm,angle=90,clip=t}
      \caption{
               Energy resolved time histories of August 24 and September 16 bursts.
               The dashed curves are exponential decay fits of the profiles. 
               In the lower panels the time histories of the blackbody radii are also
	       displayed.
              }
         \label{Fig3*}%
    \end{figure*}
%_______________________________________________________________

\section{Discussion}

On the basis of their spectral and timing properties, we interpret the three bursts detected 
from GRS 1741.9$-$2853 as type-I X-ray bursts, typically associated to low-mass binary (LMXB) systems 
(see \cite{Lewi95} for a review). 
The blackbody emission and the measured color temperatures of $\sim 2$ keV are consistent with 
this hypothesis. Spectral softening is observed in the time resolved spectra of the bursts (Table 2).
Moreover, the bursts time profiles can be fitted with exponential decays whose 
characteristic times are energy dependent, being shorter at higher energies (see Fig. 3). 
Type-I bursts strongly suggest a neutron star nature for the binary system. This indicates
GRS 1741.9$-$2853 is a transient neutron-star LMXB.

The photospheric radius expansion derived from the time resolved spectral analysis of the brightest 
burst (burst 3) can be interpreted as adiabatic expansion during an high luminosity (super-Eddington) 
type-I burst.  Actually, the 7--28 keV time history of the September 16 burst (Fig. 3, right panel)
shows a top-flattened and perhaps double-peaked profile which is typical of super-Eddington events
(e.g. \cite{Lewi95}).
Eddington-luminosity X-ray bursts can lead to an estimate of the source distance.  
Assuming a $2\times 10^{38} {\rm erg~s}^{-1}$ Eddington bolometric luminosity for a $1.4~{\rm M}_{\odot}$ 
neutron star, and taking into account the observed peak flux of burst 3 which extrapolates to an 
unabsorbed bolometric luminosity of $527\pm 42$ mCrab ($3.26\pm 0.26 \times 10^{-8} 
{\rm erg cm}^{-2}{\rm s}^{-1}$), we obtain $d=7.2\pm 0.6$ kpc.  
If we adopt the average luminosity of super-Eddington bursts proposed by Lewin, van Paradijs, \& Taam (1995) 
($3.0\pm 0.6\times 10^{38}{\rm erg s}^{-1}$) the distance value becomes $d=8.8\pm 1.2$ kpc, 
indicating GRS 1741.9$-$2853 to be very close to the Galactic Centre.
Assuming a Crab-like spectrum, for the source bolometric luminosity we derive an upper limit of 
$\sim1.6\times 10^{36}{\rm erg~s}^{-1}$ during the bursting activity (August-September 1996). 
We also obtain an average radius of $\sim6$ km for the blackbody emitting region during the bursts, 
a value supporting the neutron-star nature of the collapsed object. \\
Taking into account the intensity and the spectrum observed in the 1990 outburst (\cite{Suny90}), 
we can also derive a peak bolometric luminosity of $\sim 2\times10^{36}{\rm erg~s}^{-1}$, which
extrapolates to an accretion rate of $\la 3\times10^{-10}{\rm M}_{\odot}{\rm y}^{-1}$ 
for a canonical $1.4~{\rm M}_{\odot}$ neutron star.  These values are common among low luminosity 
LMXB transients (e.g. \cite{Tana96,Chen97}).

During the past two decades, bursting activity from LMXB transients has been reported in about 10 cases 
(e.g. Rapid Burster, Aql X-1, Cen X-4, 0748$-$673, 1658$-$298, see \cite{Hoff78,Tana96}, 
Lewin et al. 1995, and references therein), thus indicating the sources to be neutron-star binaries.
Among the LMXB transients less than $\sim 50\%$ of sources ($\simeq30\%$, 
according to Chen et al. 1997, $\simeq45\%$, according to \cite{Tana96}) are neutron-star systems,
the rest being black hole (BH) binaries.  All the BH candidates in LMXB systems are transient sources. \\
The recent (1996-1999) {\em BeppoSAX}-WFC results report on several observations 
of type-I X-ray bursts in transient sources (e.g. SAX J1750.8$-$2900, SAX J1806.5$-$2215, 
SAX J1753.5$-$2349, SAX J1808.4$-$3658, RX J170930.2$-$263927, SAX J1810.8$-$2609, see 
\cite{Heis99,Uber99}).   %Bazz97,Zand98a,Zand98b,Cocc98b,Cocc99}).
Conversely, no firm LMXB BH candidate was established.
This could imply the population of black hole LMXB to be overstimated, since most of them are 
suggested as BH candidates on the basis of their spectral characteristics only. 
Actually, for only 7 out of about 40 known transient LMXB the available mass functions suggest 
BH systems (\cite{Chen97}).

\begin{acknowledgements}
  We thank the staff of the {\em BeppoSAX Science Operation Centre} and {\em Science
  Data Centre} for their help in carrying out and processing the WFC Galactic Centre
  observations. The {\em BeppoSAX} satellite is a joint Italian and Dutch program.
  M.C., A.B., L.N. and P.U. thank Agenzia Spaziale Nazionale ({\em ASI}) for grant support.
\end{acknowledgements}

\end{document}